\documentstyle[multicol,prl,aps,epsfig]{revtex}

\begin{document}

\twocolumn[\hsize\textwidth\columnwidth\hsize\csname @twocolumnfalse\endcsname

\title{The conductance of interacting nano-wires}

\draft

\author{V.\ Meden$^{1}$ and U.\ Schollw\"ock$^{2}$}
\address{$^1$Institut f\"ur Theoretische Physik, Universit\"at G\"ottingen, 
Bunsenstr.\ 9, 
D-37073 G\"ottingen, Germany \\
$^2$Max-Planck-Institut f\"ur Festk\"orperforschung,
  Heisenbergstr.\ 1, D-70569 Stuttgart, 
Germany} 

\date{\today}
\maketitle

\begin{abstract}
The conductance of one-dimensional nano-wires of interacting
electrons connected to non-interacting leads is calculated in the
linear response regime. Two different approaches are used: a 
many-body Green function technique and a 
relation to the persistent current recently proposed
based on results of the non-interacting case. The conductance  
is evaluated using the functional renormalization group method and 
the density matrix renormalization group algorithm. Our results  
give a strong indication that the idea of obtaining the conductance 
from the persistent current holds for interacting systems. 

\end{abstract}

\pacs{71.10.Pm, 73.23.Ra, 73.63.-b}

\vspace{-0.6cm}

\vskip 2pc]
\vskip 0.1 truein
\narrowtext

The last few years brought fast progress in the experimental 
techniques to design and manipulate mesoscopic quasi-one-dimensional 
electron systems.
Electron correlations play a
crucial role in one spatial dimension leading to Luttinger
liquid behavior in the infinite wire limit.\cite{Haldane1} 
Furthermore the interplay of electron correlations and impurities leads
to a drastic effect. Even in the presence of a single impurity 
in the infinite wire limit and at low energy scales physical observables 
behave as if the system is split in two chains with open boundary
conditions at the end points.\cite{KaneFisher} 
A deeper theoretical understanding of the electron transport through
interacting nano-wires is thus of great importance in connection with
experiments and possible applications in nanotechnology. Additionally 
it will shed light on the question how electron correlations and the
interplay of correlations and impurities influence the physical
behavior of finite one-dimensional systems. 

In a typical experimental setup the wire - or more general the
nano-system - is connected to leads which act as reservoirs and the
two-terminal conductance is measured. The leads are considered to be
free of impurities and noninteracting. They have different
chemical potentials $\mu_l$ and $\mu_r$ where $\mu_l-\mu_r$ is given by the
applied voltage. The wire might contain impurities and the electrons
in the wire will later be assumed to be strongly correlated. 
For non-interacting electrons, at temperature $T=0$ and in linear
response the two-terminal conductance  
is given by the Landauer formula $G=2 e^2 |T(k_F)|^2 /h $ with 
the electron charge $e$ and the transmission probability 
$|T(k_F)|^2$ of the wire at the Fermi wave vector 
$k_F$.\cite{Landau,Buettiker} The factor $2$ is related to
the two spin directions. In the following we consider spinless
fermions and use the dimensionless conductance $g= G  h/e^2$. 
The above expression was derived using
scattering theory. For an interacting wire at $T=0$ and in linear
response $g$ can be expressed by the one-particle Green
function of the wire taken at the chemical potential and 
calculated in the presence of the non-interacting semi-infinite 
leads.\cite{Langer,Oguri1} 

Unfortunately even this Green function cannot be calculated
exactly in most cases of physical interest. Thus appropriate
approximations are required. We  use a functional 
renormalization group (RG) method\cite{Polchinski,Wetterich} 
which we have applied successfully 
in the context of impurities in interacting one-dimensional 
systems.\cite{VM1,VM2} We neglect the flow of the
two-particle vertex which provides a good approximation to the
exact result for interactions which are not too 
large (see below).\cite{VM1,VM2} 
This establishes one way to approximately determine $g$.

Here we  exclusively study the zero temperature and linear 
response regime. We focus on the case of an impurity free
interacting wire. The generalization to the case including impurities
is straightforward. 
The wire will  be modeled by the lattice model of spinless 
fermions with nearest neighbor interaction $U$ and the leads by a 
spinless tight-binding Hamiltonian. 

Within Luttinger liquid theory the conductance of an impurity free
interacting wire has been determined using an effective field  theory
and bosonization.\cite{Schulz,Maslov}
Even though these works were clearly a step forward in clarifying the
issue that for ``perfectly connected'' leads the conductance is 
independent of
the electron-electron interaction and the length of the wire, 
i.e.\ $g=1$, in connection with experimentally accessible wires the 
modeling has to be doubted. 
Within the effective field theory it is clear what
``perfectly connected'' means; a sharp jump of the Luttinger liquid
parameter $K=1$ in the leads to $K< 1$ in the interacting region.
Assuming a spatially varying $K$ is an extension of standard  
Luttinger liquid theory and it is not clear how to describe such a
situation starting from a microscopic model. As later discussed 
briefly we find $g=1$ for a lattice model in which the
interaction is turned on infinitely slowly (in space). 
\cite{Molina,VM3} For the generic case 
relevant for experiments $g$ depends on the interaction and the 
length of the wire even if no additional one-particle 
scattering terms are considered at the contacts and in the wire.  

Very recently Sushkov suggested to 
calculate $g$ from the persistent current $I(\phi)$ 
of a one-dimensional ring penetrated by a magnetic flux
$\phi$.\cite{Sushkov} The
ring of $N$ lattice sites is build of two parts: the interacting wire
with $N_W$ sites, and non-interacting ``leads'' with $N_L$ sites.
The conductance for a fixed $N_W$ is then given by\cite{Sushkov,Molina} 
\begin{eqnarray} 
g=\lim_{N_L \to \infty} \left(\frac{I(\pi/2)}{I_0(\pi/2)} \right)^2 \; , 
\label{sushkovformel}
\end{eqnarray}
where $I_0(\phi)$ denotes the persistent current of a completely
non-interacting ring with $N=N_W+N_L$ lattice sites. 
The idea can only be motivated using the concepts of the
non-interacting case.\cite{footnote1} 
The leading $1/N$ behavior of the persistent
current in a non-interacting model with $N$ lattice sites and a
localized scattering potential with transmission probability $|T(k_F)|^2$
is given by\cite{Gogolin}
\begin{eqnarray}
I(\phi) = \frac{v_F}{\pi N}  \frac{\arccos{\left( |T(k_F)|
      \cos{\left[\phi -\pi \right]} \right) }}{\sqrt{1- |T(k_F)|^2
    \cos^2{\phi }}} |T(k_F)| \sin{\phi} \; . 
\label{U0current}
\end{eqnarray}
Evaluated at $\phi = \pi/2$  this gives\cite{footnote2} 
\begin{eqnarray}
I(\pi/2) = \frac{v_F}{2 N} |T(k_F)| \; .
\label{U0currentpi2}
\end{eqnarray} 
Taking into account the Landauer formula for
the non-interacting case Eq.\ (\ref{sushkovformel}) follows immediately.
If one now considers the interacting wire as a
localized impurity in the otherwise non-interacting ring\cite{Molina} 
and assumes that the persistent current as well
as the conductance are given by the same effective transmission
probability Eq.\ (\ref{sushkovformel}) is also plausible for the
interacting case.
Using non-perturbative
methods we give a strong indication, that the method indeed works
for interacting wires.
The persistent current
can be calculated from the groundstate energy $E_0(\phi)$ 
taking the derivative with respect to the flux $I(\phi) = -d
E_0(\phi)/d \phi$. In contrast to the Green function method 
following this suggestion
the conductance $g$ is accessible using the high
precision density matrix renormalization group (DMRG) 
algorithm.\cite{DMRGbasics} 
We have earlier calculated the persistent current in interacting 
rings using the RG and a complex-valued  version of the DMRG.\cite{VM2}
This gives a second approach to determine $g$, using either the essentially
exact DMRG or the approximate RG. 

We show that within the RG
both approaches lead to the same $g$ and that for not too large
interactions the indirect method (via the persistent current) leads 
to equal conductances using RG and DMRG. We thus reach a very high
consistency using different approaches and methods. This gives us
confidence that Sushkov's suggestion can indeed be used 
for interacting wires and that the functional RG approach is
applicable also to more complex nano-systems. A generalization to wires
including impurities is straightforward and within the infinite leads 
approach the conductance in the presence of multichannel leads can be
determined.\cite{VM3}  
In contrast to Ref.\ \onlinecite{Molina}, where DMRG is only used for real
Hamiltonian matrices and the phase sensitivity
$\Delta E_0=(L/2) | E_0(0) - E_0(\pi)|$ 
is considered as a measure for the persistent current at $\phi=\pi/2$ 
we are able to calculate $I(\pi/2)$. 

We consider the Hamiltonian ($N=N_W+N_L$)\cite{footnote3} 
\begin{eqnarray}
H & = & - \sum_{j=-N_L/2-1}^{N_W+N_L/2} 
\left( c_j^{\dag} c_{j+1}^{} e^{i \phi/ N}+
  c_{j+1}^{\dag} c_j^{} e^{-i \phi/ N} \right) \nonumber \\ && 
+ U \sum_{j=1}^{N_W-1} \left(n_j-1/2 \right)  \left(n_{j+1}-1/2\right) \; ,
\label{spinlessfermdef}
\end{eqnarray}
in standard second-quantized notation. The hopping matrix element and 
the lattice constant are set to one and the flux $\phi$ is measured in
units of the flux quantum. When calculating the persistent current
periodic boundary conditions are used. When using the infinite leads
method $\phi$ is set to zero. The interaction in the wire
(lattice sites 1 to $N_W$) has to be compensated by an external 
potential. Otherwise the fermions would depopulate 
the interacting region. We here exclusively consider the case of a
half-filled system. The connection between the leads and the 
wire is considered to be perfect in the sense that no local scattering 
terms, as e.g.\ a smaller hopping matrix element between the sites 
$0$ and $1$ (and $N_W$ and $N_W+1$) or site impurities on the sites 
$0$ and $N_W+1$, are included. Such terms can be added and the more complex 
Hamiltonian can still be treated by the methods applied in this 
work.

In the infinite leads limit and for
the model discussed here $g$ is given by\cite{Langer,Oguri1} 
\begin{eqnarray}
g=4 \left| G_{N_W,1}(0) \right|^2 \; ,
\label{formel}
\end{eqnarray}
where $G_{N_W,1}$  is the interacting one-particle 
Green function in the Wannier basis. 
In Refs.\ \onlinecite{VM1} and \onlinecite{VM2}  
we describe how to determine Green functions using the functional RG
method of a system with interaction on all sites. 
As in these investigations we 
neglect the flow of the two-particle vertex. In the above references 
it was demonstrated that this approximation is valid for interactions
as large as $U \approx 1$. Later it will be shown that the same
holds here. Crucially within the approximation physical observables 
still show typical Luttinger liquid behavior (for $U \leq 2$) 
in the infinite wire limit.\cite{VM1,VM2} Thus the relevant 
correlations are included. This has to be contrasted to simple 
perturbative methods \cite{Sushkov,Oguri} which do not include 
Luttinger liquid effects. In the present context the
infinite non-interacting leads can be integrated out only leading to 
additional contributions on the lattice sites $1$ and $N_W$. The effective
Hamiltonian can then be treated using the RG and  
$G_{N_W,1}(\varepsilon)$ can be calculated numerically. 
The scattering properties can be understood in
terms of an effective oscillating one-particle potential. This 
helps to obtain an intuitive understanding of the observed 
physics.\cite{VM1,VM2} Technical details about this method will be presented 
elsewhere.\cite{VM3} 

\begin{figure}[hbt]
\begin{center}
\vspace{0.5cm}
\leavevmode
\epsfxsize7.4cm
\epsffile{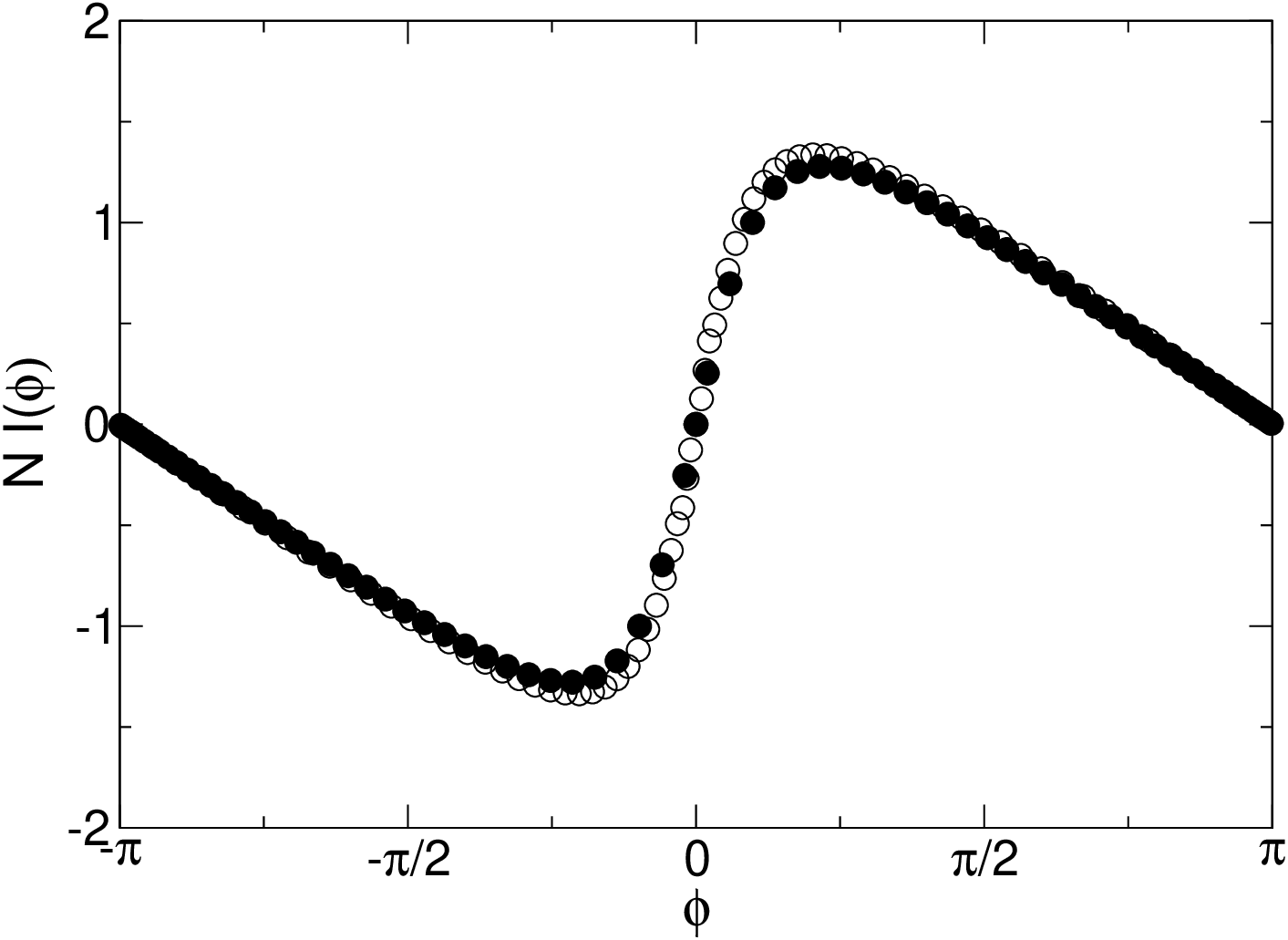}
\vspace{0.4cm}
\caption{Persistent current $N I(\phi)$ for $N=64$, $N_W=12$, and
  $U=1$. The filled symbols are DMRG data and the open symbols RG data.}
\label{fig1}
\vspace{-0.4cm}
\end{center}
\end{figure}

\begin{figure}[hbt]
\begin{center}
\vspace{0.5cm}
\leavevmode
\epsfxsize7.4cm
\epsffile{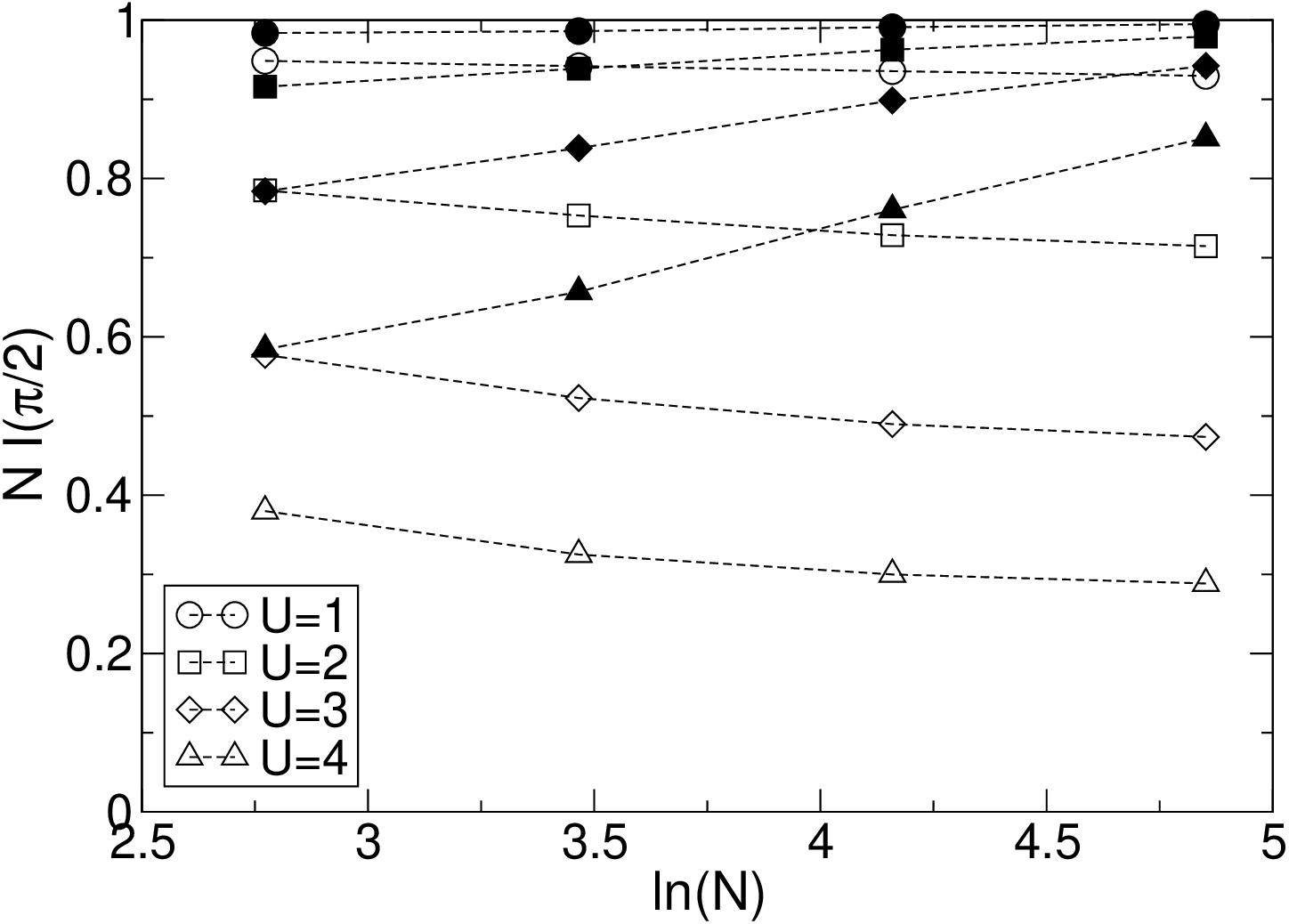}
\vspace{0.4cm}
\caption{$N I(\pi/2)$ for $N_W=12$ (open symbols), $N_W=13$
  (filled symbols), different $N$, and $U$ obtained from DMRG.}
\label{fig2}
\vspace{-0.4cm}
\end{center}
\end{figure}

In Ref.\ \onlinecite{VM2} we have
investigated the persistent current of a ring with interaction on all
$N$ sites using
RG and complex-valued DMRG. A generalization to the present situation is
straightforward. Fig.\ \ref{fig1} shows typical curves for 
$N I(\phi)$ at even $N_W$  obtained from DMRG
and RG. The parameters are $N=64$, $N_W=12$, and $U=1$.
Both curves agree quantitatively. This holds for even 
larger interactions (see also Ref.\ \onlinecite{VM2}) 
and for all $N$ and even $N_W$ we have investigated. 
The current for this typical
set of parameters is far from being sinusoidal. Thus the phase
sensitivity $\Delta E_0$ can obviously not be used directly 
to determine $I(\pi/2)$ and $g$. 
Only in the limit of a strong effective impurity
$I(\phi)$ becomes proportional to $\sin{\phi}$.\cite{VM2} 
In the present context and for even $N_W$ 
this limit is reached for very large $U$ and $N_W$. 
Increasing $N$ for fixed $N_W$ (even) the curves of Fig.\ \ref{fig1}
only change slightly and the
infinite leads limit can reliably be extrapolated  from $N \approx
100$.
The $N I(\phi)$ extrapolated to the infinite lead limit 
decreases if $N_W$ (even) is increased. This
is related to the Luttinger liquid scaling of $I(\phi)$.\cite{VM2,Gogolin}
For odd $N_W$ and all $U$ we have studied 
the persistent current is of almost saw tooth like shape. Increasing
$N$ at fixed $N_W$ the data seem to converge to the non-interacting
and impurity free curve $ N I_0(\phi) = v_F (\phi-\pi)/ \pi$, 
with $0 \leq \phi < 2 \pi$ and the Fermi velocity $v_F$,\cite{VM2} 
which implies $g=1$.
To obtain the conductance following the suggestion of Sushkov we have
determined $E_0(\phi)$ (and the non-interacting groundstate energy
$E_0^0(\phi)$) for a few fluxes around $\pi/2$ using DMRG and RG. 
From this $I(\pi/2)$ and 
$I_0(\pi/2)$ were obtained by numerical
differentiation. Since around $\phi = \pi/2$, $E_0(\phi)$ is a fairly
smooth function (see 
 also Fig.\ \ref{fig1}), 
this procedure leads to reliable results.
For fixed $N_W$ and $U$ we have done this for $N=16$, 
$32$, $64$, 
and $128$. As an example in Fig.\ \ref{fig2} we present DMRG data of 
$N I(\pi/2)$ for $N_W=12$ and $N_W=13$ and different $U$ as a function
of $\ln{N}$. To extrapolate we have fitted the data to a quadratic 
polynomial in $1/N$. For even $N_W$ the $N \to \infty$ limit of $N I(\pi/2)$
can be obtained up to high precision.  
We have also tried different extrapolation
schemes to ensure that the asymptotic current only very weakly depends 
on the details of the assumed scaling.
For odd $N_W$ and $U\geq 3$ the finite $N_W$ data still change 
significantly even for $N$ as large as $128$ and the asymptotic 
limit obtained by extrapolation is clearly less reliable.
The DMRG data strongly tend to $N I(\pi/2)=1$ which corresponds to $g=1$. 
The same behavior can be found for other odd $N_W$. This suggests  
$g=1$ for all odd $N_W$ but a definite statement would require larger $N$.
The RG data for $N I(\pi/2)$ show a quite similar behavior, with the
difference that using this technique for odd $N_W$ the large $N$ data
are closer to 1 and clearly extrapolate to $g=1$. 
Additionally $g$ was determined within the RG using Eq.\
(\ref{formel}) which already implies infinite leads.

\begin{figure}[hbt]
\begin{center}
\vspace{0.6cm}
\leavevmode
\epsfxsize7.4cm
\epsffile{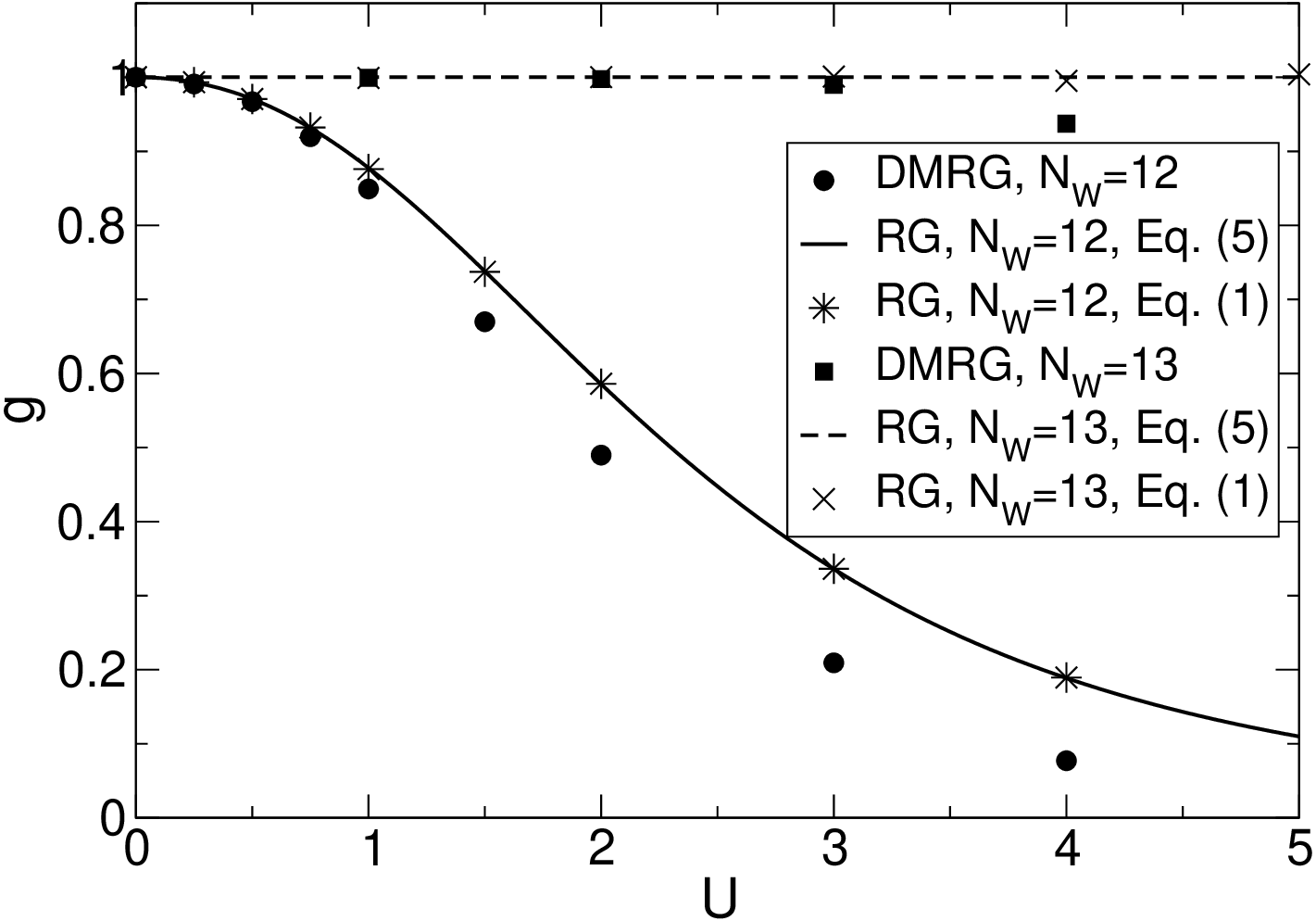}
\vspace{0.4cm}
\caption{Dimensionless conductance $g$ as a function of $U$ for
  $N_W=12$ and $N_W=13$.}
\label{fig3}
\vspace{-0.4cm}
\end{center}
\end{figure}

Fig.\ \ref{fig3} summarizes our results for $N_W=12$ and $N_W=13$,
which are typical also for other $N_W$.
Several conclusions can be drawn from this figure. Within
the RG the two approaches to determine $g$ lead to results which agree
quite well. The small deviation between the two data sets has mainly 
two sources, both of them making the persistent current data 
[Eq.\ (\ref{sushkovformel})] less precise than the infinite leads data
[Eq.\ (\ref{formel})]:
(i) The numerical differentiation of $E_0(\phi)$. (ii) 
The $N_L \to \infty$  extrapolation. 
The estimated total error is smaller than the symbol size.
Fig.\ \ref{fig3} shows that Sushkov's suggestion Eq.\ (\ref{sushkovformel})
indeed works for an interacting wire. 
For odd $N_W$ in both approaches we find perfect conductance $g=1$ for
all $U$.  
The infinite leads approach does not require an extrapolation
and  solving the flow equations within 
this approach is much faster compared to calculating the persistent
current in a $N$ lattice site system. Thus within the RG and 
also in perturbation theory\cite{Oguri} using Eq.\
(\ref{formel}) is clearly superior to the persistent current
approach. The advantage of the persistent current
method is that it enables us to determine $g$ using DMRG. 
For up to $U \approx 1$ the RG and DMRG data agree quantitatively. For
larger $U$ the RG data still show the same qualitative behavior as the
DMRG results. As discussed in connection with Fig.\ \ref{fig2} 
the $N \to \infty$ extrapolation for odd $N_W$ and 
large $U \geq 3$ leads to results which
are less reliable. 
The surprising even-odd effect, observable also at other $N_W$,
has been discussed earlier within the Hubbard model using
perturbation theory.\cite{Oguri} 
Similar to perturbation theory within the RG the effect can
be traced back to the combination of particle-hole and site inversion
symmetry.\cite{VM3} Using the RG the interaction generates a modulation  
of the hopping matrix element between the sites of the interacting 
wire. This acts as an effective impurity which for odd $N_W$ 
has a perfect resonance at $k_F$. 
For even $N_W$ and small $U$
the conductance goes like $g=1-c(N_W) U^2$ with a $c(N_W) > 0 $
which increases with increasing $N_W$.

For fixed $U$, $g$ is a decreasing function of $N_W$ (even). In the
limit of very large $N_W$ we find\cite{VM3} a power-law suppression  
$g(U,N_W) \sim N_W^{2(1-1/K)}$ known from the problem of a single
impurity in a Luttinger liquid.\cite{KaneFisher} Here the 
``switching on'' of the interaction acts as an impurity. This behavior 
cannot be captured by simple perturbation theory\cite{Oguri} since 
it requires Luttinger liquid effects to be included. In the 
artificial limit of an interaction turned on infinitely 
slowly (in space) this suppression can be 
avoided leading to $g(U,N_W)=1$ also for even $N_W$.\cite{Molina,VM3}     

In this work, we determine the conductance of an interacting 
nano-system (i) by a Green function technique and (ii) using the 
mapping on a persistent current problem suggested by 
Sushkov,\cite{Sushkov,Molina} both at the level of a functional RG 
calculation. The resulting conductances are equal and show 
for interactions $U \leq 1$ (iii) excellent agreement with the 
conductance obtained following Sushkov's idea and using essentially 
exact complex-valued DMRG calculations of the numerical key quantity, 
the persistent current $I(\pi/2)$. 
(i)-(iii) give a very strong indication that Sushkov's suggestion,
which analytically can only be justified for non-interacting
nano-systems also holds for the interacting case. 
The conductance can be obtained for
microscopic models avoiding the mapping to an effective field theory
using bosonization. 
As an application we calculate the
conductance of an interacting, impurity free nano-wire
connected to infinite leads. The methods used can easily be
generalized to (i) more complex nano-systems with different geometries, 
(ii) wires with interaction and impurities, (iii) more complex 
contacts, (iv) multi-channel leads, (v) models with spin, and (vi) 
other filling factors. 
We thus believe that combining the two approaches (infinite lead,
persistent current) and the two methods (DMRG, RG) gives a very
powerful tool to investigate transport properties of realistic models 
for complex interacting nano-systems. 
 
We thank K.\ Sch\"onhammer, W.\ Metzner, H.\ Schoeller, and W.\
Zwerger for very valuable discussions.
U.S.\ is grateful to the Deutsche Forschungsgemeinschaft 
for support from the Gerhard-Hess-Preis and V.M.\ acknowledges 
support from the Bundesministerium f\"ur Bildung und Forschung 
(Juniorprofessor Programm).



\begin{thebibliography}{*}
\bibitem{footnote1} We here present a ``derivation'' different from
  Ref.\ \onlinecite{Sushkov}.
\bibitem{footnote2} In contrast to a statement in 
Ref.\ \onlinecite{Sushkov} the current
is maximal at $\phi=\pi/2$ only in the limit of a very
strong impurity.
\bibitem{footnote3} Without loss of generality we only consider even
  $N_L$.
\end{thebibliography}
\end{document}